
\input phyzzx

\def\IMPA#1{{\sl Int. J. Mod. Phys. {\bf A#1}}}

\def\NPB#1{{\sl Nucl.\ Phys.\ {\bf B#1}}}

\def\PLB#1{{\sl Phys.\ Lett.\ {\bf #1B}}}
\def\PRep#1{{\sl Phys.\ Rep.\ {\bf #1}}}

\def\PRD#1{{\sl Phys.\ Rev.\   {\bf D #1}}}
\def\PRL#1{{\sl Phys.\ Rev.\ Lett.\ {\bf #1}}}


\def\nxl{\hfill\break}



\def\H{{\cal H}}                            

\def\a{\alpha}

\def\e{\epsilon}

\def\s{\sigma}


\def\o{\over}

\def\bold#1{\setbox0=\hbox{$#1$}
     \kern-.025em\copy0\kern-\wd0
     \kern.05em\copy0\kern-\wd0
     \kern-.025em\raise.0433em\box0 }
\def\lowmp{\lower.11em\hbox{${\scriptstyle\mp}$}}

\def\frac#1#2{{\textstyle{
 #1 \over #2 }}}                            


\def\1{{\rm 1 \!\!\, l}}                        
%

%
%


\hyphenation{Di-par-ti-men-to}
\hyphenation{na-me-ly}
\hyphenation{al-go-ri-thm}
\hyphenation{pre-ci-sion}
\hyphenation{cal-cu-la-ted}

%

%

\Pubnum={$\rm PAR\; LPTHE\; 92/32;\;  hep-th / 9209047$}
\date={}
\titlepage
\title{EXACT STRING SOLUTIONS IN 2+1-DIMENSIONAL DE SITTER SPACETIME}
\author{ H.J. de Vega }
\address{ Laboratoire de Physique Th\'eorique et Hautes Energies
     \foot{Laboratoire Associ\'e au CNRS UA 280 \nxl
     Postal address: \nxl
           L.P.T.H.E., Tour 16, $1^{\rm er}$ \'etage,
Universit\'e Paris VI,\nxl
           4, Place Jussieu, 75252, Paris cedex 05, FRANCE. }, Paris}
\author{A. V. Mikhailov }
\address{Landau Institute for Theoretical Physics,  Russian Academy of
Sciences\foot{ Ul. Kossyguina 2, 117334 Moscow, RUSSIA}
and Observatoire de Paris, Section de Meudon, Demirm$^{\ast}$ }
\author{N. S\'anchez}
\address{Observatoire de Paris, Section de Meudon, Demirm
\foot{Laboratoire Associ\'e au CNRS UA 336, Observatoire de Meudon et
\'Ecole Normale Sup\'erieure. \nxl Postal address: \nxl
DEMIRM, Observatoire de Paris. Section de Meudon, 92195 MEUDON
Principal Cedex, FRANCE.}}
\endpage
\vfil
\abstract
\vskip 1cm
Exact and explicit string solutions in
de Sitter spacetime are found.
(Here, the string equations reduce to a sinh-Gordon model).
A new feature without flat spacetime analogy appears:
starting with a single world-sheet, several
(here two) strings emerge. One string is stable
and the other (unstable) grows as the universe grows.
Their invariant size and energy either grow as the
expansion factor or tend to  constant.
Moreover, strings can  expand (contract) for large (small) universe radius
with a different rate than the universe.

\medskip
\endpage
\vskip 1cm
\sequentialequations
\REF\erice{H.J.~de~Vega and N. S\'anchez, Lectures delivered
at the Erice School ``String Quantum Gravity and Physics at the Planck
scale'', 21-28 June 1992, Proceedings edited by N. S\'anchez,
World Scientific, 1993.}
\REF\twbk{See for a review, T.W.B. Kibble, Erice Lectures at the
Chalonge School in Astrofundamental Physics,
N. S\'anchez editor, World Scientific, 1992.}
\REF\desi{H. J. de Vega and N. S\'anchez, \PRD{47}, 3394 (1993).}
\REF\mult{F. Combes, H.J.~de~Vega, A.V.~Mikhailov and N.~S\'anchez,
lpthe preprint 93/44. }
\REF\nos{H. J. de Vega and N. S\'anchez,  \PLB{197}, 320 (1987).}
\REF\vene{N. S\'anchez and G. Veneziano, \NPB{333}, 253 (1990), \nxl
M. Gasperini, N.S\'anchez and G. Veneziano, \nxl
\IMPA{6}, 3853 (1991) and \NPB{364}, 365 (1991).}
\REF\vil{ A. Vilenkin, \PRD{24},  2082 (1981),
 \PRep{121}, 263 (1985) .\nxl
N. Turok and P. Bhattacharjee, \PRD{29} 1557 (1984).}
\REF\ani{ H. J. de Vega and I. L. Egusquiza, LPTHE preprint 93/43.}
\REF\tuba{ N. Turok, \PRL{60}, 549 (1988). \nxl
J. D. Barrow, \NPB{310}, 743 (1988). }

String propagation in curved spacetimes reveals new insights for
string theory.
The string behaviour in strong gravitational fields and in the
vicinity of spacetime singularities is especially interesting.
In gravitational shock-waves and singular plane wave  spacetimes
the string equations are exactly (and explicitly)
solvable even at the spacetime singularities uncovering a rich variety
of physical phenomena[\erice ].
The classical string propagation in D-dimensional de Sitter spacetime
 is an
integrable model as it has been explicitly shown in ref.[\desi] where, in
addition, all the 1+1
dimensional string solutions were given. It is clearly an
appealing challenge to explicitly solve the string
propagation in such a relevant cosmological spacetime as
the de Sitter universe.

The string solutions reported here indeed apply to cosmic strings
in de Sitter spacetime as well. The dynamics of cosmic strings
in expanding universes has been studied in the literature for the
Friedman-Robertson-Walker (FRW)  cases (see for example [\twbk ,
 \vil , \ani ]). It must be noticed that the string behaviour we found
here in de Sitter universe is essentially {\bf different}
from the standard FRW where $R(t_0)$ is a positive power of the cosmic
time $t_0$. In such FRW universes, strings always oscillate in time,
 the comoving spatial string coordinates {\it contract} and the proper
string size stays {\it constant} asymptotically for  $t_0 \to \infty$.
In the cosmic string literature this is known as 'string stretching'.
We called such behaviour 'stable' [\vene ]. On the contrary, in
de Sitter spacetime, as we show below, two
types of asymptotic behaviors are present : (i) the proper string size
and energy grow with the expansion factor ('unstable' behaviour) or
(ii) they tend to constant values ('stable' strings).

The unstable string solutions in de Sitter universe may provide a
mechanism to self-sustain inflation as proposed in refs.[\tuba]-[\vene
] without advocating an inflaton field. The present note is a first
step in the investigation of multi-string exact solutions in
de Sitter spacetime using soliton methods [\mult ]. Such general solutions
should provide essential clues about the feasability of inflationary
string scenarios.

In this letter we present exact and explicit solutions for strings
propagating in the 2+1 dimensional de Sitter spacetime.
In this case, de Sitter spacetime can be considered as a 3-dimensional
hyperboloid embedded in a 4 dimensional flat Minkowski spacetime with
coordinates $(q_0,q_1,q_2,q_3)$ and metric
$$
  ds^{2} = {1\o{H^2}} [ - (dq^0)^2 +(dq^1)^2+(dq^2)^2+(dq^3)^2 ]			\eqn\metrica
$$
where H is Hubble's constant and
$$
	(q^0)^2 = (q^1)^2 + (q^2)^2 + (q^3)^2  - 1 	\quad ,
							\eqn\hiperb
$$
In the comoving coordinates $(t_0, X^1 , X^2 )$ the de Sitter metric
takes the form
$$
ds^{2} = -(dt_0)^2 + e^{2 H t_0}~\left[\, (d X^1)^2 + (d X^2)^2 \, \right]
\eqn\comov
$$
where the cosmic time $t_0$ and the conformal time $\eta$ are given by
$$
\eta = - {1\o{H}}\exp[-Ht_0] = -{1\o{H(q_0 + q_1)}}
					\eqn\tiempos
$$
and
$$
X^1 = {{q^2}\over{H(q^0 + q^1)}}~~,~~X^2 = {{q^3}\over{H(q^0 + q^1)}}~~,
\eqn\equis
$$
The string equations of motion take here the form
$$
{{\partial^2{q}}\o{{\partial{x_-}\partial{x_+}}}} +
 [{{\partial{q}}\o{\partial x_+}}.{\partial q \o{\partial x_-}}]\;q  = 0
\eqn\movim
$$
where . stands for the Lorentzian scalar product $a.b \equiv -a_0b_0 +
a_1b_1 + a_2b_2 + a_3b_3$  , $x_\pm \equiv {1\o{2}}(\tau \pm \s)$ ,
$\s$ and $\tau$ being the string world sheet coordinates.
 In addition, we have eq.\hiperb\ , i.e.  $q.q = 1$ ,
 and the string constraints on the world sheet are
$$
 T_{\pm\pm}= {\partial q \o {\partial x_\pm}}.{\partial q \o {\partial x_\pm}}
= 0				\eqn\vincu
$$
We define
$$
\exp[\a(\s,\tau)] = -{\partial q \o { \partial x_-}}  . {\partial q \o
{ \partial x_+}}
			\eqn\alfa
$$
As it is shown in ref.[\desi], $\a(\s,\tau)$ obeys the sh-Gordon equation
$$
 {{\partial^2 \a}\o{\partial \tau ^2 }} - {{\partial^2 \a}\o{\partial
\s ^2}} -\exp\a + \exp-\a  = 0
						\eqn\shGordon
$$
Notice that for closed strings $q(\s,\tau)$ and hence $\a(\s,\tau)$
are periodic functions of $\s$ with period $2\pi$.
Therefore, to find string solutions in de Sitter spacetime we can start
from a periodic solution of eq.\shGordon , and insert it on the field
equations \movim :
$$\eqalign{
[{{\partial^2 }\o{\partial \tau ^ 2}} -
{{\partial^2 }\o{\partial \s ^2}}
 - \exp\a(\sigma,\tau)\;]q(\sigma,\tau)=0 \cr}
							\eqn\kgg
$$
Once this {\bf linear} equation in $q(\sigma,\tau)$ is solved, it
remains to impose the constraints \vincu\ and eq.\alfa.

Let us remark that  $\exp[\a(\s,\tau)]$  has a clear physical
interpretation. The invariant interval between two points on the
string computed with the spacetime metric \metrica\ is given by
$$
ds^2 ={1\o{H^2}}dq.dq = {1\o{2H^2}} \; \exp[\a(\s,\tau)]\;(d\s^2-d\tau^2)
							\eqn\intcuer
$$
Therefore we can define
$$
S(\s, \tau) \equiv {1 \o {\sqrt2 H}}\exp[\a(\s,\tau)/2]
\eqn\tama
$$
as the invariant string size.
The energy density for the sinh-Gordon model reads here
$$
\H = {1\o2}[({{\partial \alpha}\o{\partial \tau}})^2+({{\partial
\alpha}\o{\partial \s}})^2] - 2\cosh\a
					\eqn\energia
$$
That means a potential unbounded from below
$$
V_{eff} = - 2 \cosh\a   \quad  ,
					\eqn\potencial
$$
with absolute minima at $\a = +\infty$ and at $\a = -\infty$.
As the time $\tau$ evolves, $\a(\s,\tau)$ will generically approach
these infinite minima. The first minimum corresponds to an infinitely large
string whereas the second describes a collapsed situation. That means
that strings in de Sitter spacetime will generically tend either to
inflate at the same rate as the universe (when $\a\to+\infty$) or to
collapse to a point (when $\a\to-\infty$).
As we shall see below these general trends are confirmed by the
explicit string solutions.
Let us start by studying solutions where  $\a=\a(\tau)$. Then, the
energy is
$$
{1\o2}{\alpha^{\prime}}^2  - 2 \cosh\a = E = {\rm constant}
				\eqn\enerII
$$
$\a(\tau)$ describes the position of a non-relativistic particle with
unit mass rolling down the effective potential $V_{eff}=-2\cosh\a$ . A
particularly interesting situation is the critical case $E = -2$ when
one starts to roll down from the maximun of $V_{eff}$. That is, the
initial speed is zero and the 'time' $\tau$ to reach either minimun
($\a=\infty \;  or \;  -\infty$) is infinity. The corresponding solutions are
$$
\a_-(\tau)= \log\left[\coth^2 \left({{\tau}\o{\sqrt{2}}}\right)\right]
\quad   {\rm and} \quad
\a_+(\tau)= \log\left[\tanh^2 \left({{\tau}\o{\sqrt{2}}}\right)\right]
				\eqn\solualfa
$$
$\a_-(\tau)$ starts at $\a=0$ for $\tau=-\infty$ and rolls down to the
{\bf right} reaching $\a=+\infty$ for $\tau\to 0^-$. The behaviour of
$\a_-(\tau)$ near the initial and final points is as follows:
$$\eqalign{
\a_-(\tau)\quad \buildrel{\tau\to-\infty}\over= \; 4 \, e^{\tau \sqrt{2}} \; +
\;
O(e^{2 \tau \sqrt{2}})  \cr
\a_-(\tau)\quad \buildrel{\tau\to0}\over= \log{{2}\o{\tau^2}}
+{1\o{3}}\tau^2 +
O(\tau^4) \cr}
				\eqn\amenasin
$$
The solution $\a_+(\tau)$ also starts at $\a = 0$ for $\tau = -\infty$
but rolls down to the {\bf left} reaching $\a=-\infty$ for $\tau \to 0^-$.
We have for $\a_+(\tau)$ :
$$\eqalign{
\a_+(\tau)\quad \buildrel{\tau\to-\infty}\over= \; -4 \, e^{\tau \sqrt{2}}\; +
\; O(e^{2 \tau \sqrt{2}})\cr
\a_+(\tau) \quad \buildrel{\tau\to0}\over= -\log{{2}\o{\tau^2}}
-{1\o{3}}\tau^2 +\;
O(\tau^4) \cr}
				\eqn\amasasin
$$
Notice that $\a_+(\tau) = -\a_-(\tau)$. In addition we have the
trivial (but exact) solution $\a^{(o)}(\tau)\equiv0$.
Now that the function $\a(\tau)$ is known, we proceed to solve
eq.\kgg\   for $q(\s,\tau)$ with the constraints \vincu\ and \alfa\ .
 Since $q_0$ is a time-like coordinate, we shall assume $q_0 =
q_0(\tau)$.
A natural ansatz is then
$$
q=(q_0(\tau),q_1(\tau), f(\tau) \cos\s, f(\tau) \sin\s )
						\eqn\ansatzI
$$
 Then, eqs.\hiperb\ and \movim\ - \alfa\  require
$$
q_0(\tau)^2 = q_1(\tau)^2 + f(\tau)^2
				\eqn\ufaI
$$
$$
[{{dq_0}\o{d\tau}}]^2=[{{dq_1}\o{d\tau}}]^2 + [{{df}\o{d\tau}}]^2 + f^2
				\eqn\ufaII
$$
$$
e^{\a(\tau)}=[{{dq_0}\o{d\tau}}]^2-[{{dq_1}\o{d\tau}}]^2 - [{{df}\o{d\tau}}]^2
+ f^2
				\eqn\ufaIII
$$
and
$$\eqalign{
{{d^2}\o{d^2\tau}}q_0 - e^{\a(\tau)}q_0(\tau)=0 \cr
{{d^2}\o{d^2\tau}}q_1 - e^{\a(\tau)}q_1(\tau)=0 \cr
{{d^2}\o{d^2\tau}}f(\tau) + f(\tau)-e^{\a(\tau)}f(\tau)=0 \cr}
				\eqn\ansat
$$
In addition, it seems reasonable to choose the time coordinate
$q_0(\tau)$ to be an odd function of $\tau$. Remarkably enough,
eqs.\ufaI\ - \ansat\ admit consistent solutions for $\a(\tau) =
\a_+(\tau)$, $\a(\tau) = \a_-(\tau)$ and $\a(\tau) = 0$. For
$\a(\tau) = \a^{(o)}(\tau) = 0$ , we find
$$
q^{(o)}(\s,\tau)= {1\o{\sqrt{2}}}(\sinh\tau,\cosh\tau,\cos\s,\sin\s)
					\eqn\solI
$$
For $\a(\tau) = \a_-(\tau)$, we have
$$\eqalign{
q_-(\s,\tau) =
(\sinh\tau -{1\o{\sqrt{2}}}\cosh\tau \,
\coth[{1\o{\sqrt{2}}}\tau],\quad
\cosh\tau -{1\o{\sqrt{2}}}\sinh\tau \,
\coth[{1\o{\sqrt{2}}}\tau],\cr
{1\o{\sqrt{2}}}\cos\s \,
\coth[{1\o{\sqrt{2}}}\tau],\quad {1\o{\sqrt{2}}}\sin\s \,
\coth[{1\o{\sqrt{2}}}\tau]) , \qquad  \qquad \cr}
					\eqn\solII
$$
And for $\a(\tau) = \a_+(\tau)$ we find
$$\eqalign{
q_+(\s,\tau)=
(\sinh\tau -{1\o{\sqrt{2}}}\cosh\tau
\tanh[{1\o{\sqrt{2}}}\tau],\quad
\cosh\tau -{1\o{\sqrt{2}}}\sinh\tau
\tanh[{1\o{\sqrt{2}}}\tau],\cr
{1\o{\sqrt{2}}}\cos\s
\tanh[{1\o{\sqrt{2}}}\tau], \quad {1\o{\sqrt{2}}}\sin\s
\tanh[{1\o{\sqrt{2}}}\tau]) \qquad \qquad . \cr}
					  \eqn\solIII
$$
 These string solutions are given for a fixed de Sitter frame.
Applying the de Sitter group to them yields a multi-parameter family of
solutions. As it is clear, we can study them in the frame
corresponding to eqs.\solI\ - \solIII\ without loss of generality.
Let us now discuss the physical interpretation of these solutions.

The string energy can be easily computed from the spacetime
string energy-momentum tensor:
$$
\sqrt{-G}~ T^{AB}(X) = {1 \o {2\pi \a'}} \int d\s d\tau
\left( {\dot X}^A {\dot X}^B -X'^A X'^B \right) \delta^{(D)}(X - X(\s, \tau) )
\eqn\tens
$$
Whenever $t_0= t_0(\tau)$, the string energy at a time
$t_0$ is given by:
$$
E(t_0) = \int d^{D-1}X \sqrt{-G}~ T^{00}(X) = {1 \o {\a'}}
{{dt_0} \o {d\tau}}
\eqn\eneg
$$
where $\a'$ stands for the string tension.

We recall that for a given time  $ q_0 = q_0(\tau) $, the de Sitter
space is a sphere $ S^2$ with radius $R(\tau) = {1\o{H}}\sqrt{1+q_0(\tau)^2}$.
The solution $q^{(o)}(\s,\tau)$ [eq.\solI\ ], describes
a string of constant size $ S^{(o)} = {1 \o {\sqrt2 H}} $ and
constant energy $E^{(o)} = {1 \o {\a' H}} $ , in a de Sitter
universe that inflates for $\tau\to\infty$ since for this solution
$  R(\tau) = {1\o{H}}\sqrt{1+{\sinh^2 \tau \o{2}}} $.
This solution is probably unstable under small perturbations.

Let us analyze now the solution $q_-(\s,\tau)$.
Fig. 1 depicts the time coordinate $q_-(\s,\tau)_0$.
We see that this solution describes actually {\bf two strings},
since for a given value of $q_0$ , there are two values of $\tau$.
That is, $\tau$ is a two-valued function of  $q_0$. Each branch of
$\tau$ as a function of $q_0$ (or $t_0$ ) corresponds to a different
string. This an entirely new feature for strings in curved
spacetime. It has no analogy in flat spacetime where the time
coordinate obeys the D'Alambert equation and therefore one can
always choose a gauge where the time  is  proportional to $\tau$ .
The appearence of multiple strings is a generic feature
in de Sitter spacetime as shown in ref.[\mult ], where exact
{\it multistring} solutions are constructed.
For this solution $q_-(\s,\tau)$, the string size and energy are
$$
S_-(\tau) = {1\o{\sqrt{2}H}}\coth\left|{1\o{\sqrt{2}}}\tau\right|~,~~
E_-(\tau) = {1\o{\a' H}}\left| 1 +
{1 \o{\cosh(\sqrt2 \tau) - {1 \o \sqrt2} \sinh(\sqrt2 \tau) -1}}
\right| .
\eqn\talla
$$
That is, the string size increases for $\tau < 0$ and decreases for $
\tau > 0$ with a singular behaviour ${1\o{|\tau}|}$ for $\tau \to 0$.

We first analyze the inflationary expansion phase $q_0 > 0$.
For $q_0 = 0$ , the two strings correspond to (we call them I and II) :
$$\eqalign{
{\rm String}(I): \quad \tau =& -\tau_0 \cr
{\rm String} (II): \quad \tau =& +\tau_0 \cr}
$$
where $\tau_0 = 1.4890...$ is the positive root of $q_{-0}(\tau) = 0$
(see eq.\solII ). Both strings have at $q_0 = 0$ the same size
$$
S_-(\pm\tau_0) = {.903..\o H}
$$
and different energies, $E_-(\tau_0) = {4.260...\o {\a' H}} >
E_-(-\tau_0) = {1.166...\o {\a' H}} $.

For $q_0 \to +\infty $, the two strings correspond to
$$\eqalign{
{\rm String}(I) :\quad \tau =& -{1\o q_0} \to 0^- \cr
{\rm String}(II) : \quad \tau =& \log q_0 - \log\left[{{1\o2}
\left(1 - {1 \o \sqrt2} \right)}\right]
\to +\infty  \cr}
$$
Asymptotically, the string sizes and energies are
$$\eqalign{
(I):\quad S_- \buildrel{q_0 \to \infty}\over=& {1\o {H |\tau|}}
\simeq {{R(q_0)}\o H} \to +\infty \cr
 E_- \buildrel{q_0 \to \infty}\over=& {1\o {\a 'H }}
\left( {1 \o {|\tau|}} + 1 \right) \simeq {1\o {\a 'H }}
\left( R(q_0) + 1 \right) \to +\infty . \cr
(II):\quad S_- \buildrel{q_0 \to \infty}\over=& {1\o {\sqrt2 H }}~~,~~
E_- \buildrel{q_0 \to \infty}\over= {1\o {\a 'H }} \cr}
\eqn\masin
$$
We see that for the string (I) both its size and energy grow
 monotonically, this growing becoming explosive for
$q_0 \to \infty$ when the size of the de Sitter space diverges.
Actually, the string grows there at the same rate as the whole space.
This describes an {\it unstable} string.
The branch II represents a {\it stable} string for $q_0 \to \infty$,
both size and energy being asymptotically constant.
It must be noticed that the size and the energy of the string I
monotonically increase with $q_0$ whereas for the string II,
$S_-$ and $ E_-  $ both monotonically decrease with $q_0$.

Let us describe now the solution $q_-(\s,\tau)$ for $q_0 < 0$, that
is in the contracting phase of de Sitter universe. For $q_0 \to -\infty$
we have for the strings I and II :
$$\eqalign{
(I):\quad \tau =& -\log(-q_0) + \log\left[{{1\o2}
\left(1 - {1 \o \sqrt2} \right)}\right]
\to -\infty  \cr
(II): \quad \tau =&-{1\o q_0} \to 0^+ \cr}
$$
The string size and energy are in this limit:
$$\eqalign{
(I):\quad S_- \buildrel{q_0 \to -\infty}\over=& {1\o {\sqrt2 H }}~~,~~
E_- \buildrel{q_0 \to -\infty}\over= {1\o {\a 'H }} \cr
(II):\quad S_- \buildrel{q_0 \to -\infty}\over=& {1\o {H \tau}}
\simeq {{R(q_0)}\o H} \to +\infty \cr
 E_- \buildrel{q_0 \to -\infty}\over=& {1\o {\a 'H }}
\left( {1 \o {\tau}} + 1 \right) \simeq {1\o {\a 'H }}
\left( R(q_0) + 1 \right) \to +\infty . \cr}
\eqn\menin
$$
The string I starts in the contracting phase with a stable behaviour
for  $q_0 \to -\infty$, while string II starts with an infinite size
(the size of the de Sitter space) for $q_0 \to -\infty$.
The size and energy of the string I grow monotonically
with $q_0$. String II
contracts with the universe itself for  $q_0 < 0$ and continues
to contract for $q_0 > 0$ until reaches the constant value
$S_- = {1\o {\sqrt2 H }}$ for  $q_0 \to \infty$.

The behaviour for small $|\tau|$ confirms the asymptotic results found in
refs.[\desi - \nos - \vene ].

It is interesting to study this string solution in the
comoving de Sitter coordinates. The cosmic time $t_0$ and
the conformal time $\eta$
 [eq.\tiempos\ ] take for $q_-(\s,\tau)$ the form:
$$\eqalign{
e^{Ht_0} = - {1\o{H\eta}}= \left|1 - {1\o{\sqrt{2}}}
\coth({1\o{\sqrt{2}}}\tau)\right| \; e^{\tau} \cr
\rho = {{e^{-\tau}}\o{\left|1 - \sqrt{2}
\tanh({1\o{\sqrt{2}}}\tau)\right|}} \qquad
   	\cr}	\eqn\coorden
$$
where
$$ \rho \equiv \sqrt{ (X^1)^2 + (X^2)^2 }
					\eqn\ro
$$
Therefore, for $t_0 \to +\infty$, we have :
$$\eqalign{
{\rm string}(I):\quad\eta =& {{\tau}\o{H}}\to 0^- ~,   \qquad
\rho = {1\o H} + O(\tau^2) \cr
t_0 =& -{1\o{H}}\log|\tau| + O(\tau) \to +\infty \cr
{\rm string}(II):\quad\eta =& -{{2 + \sqrt2}\o{H}}\;e^{-\tau} \to 0^-
{~},~ \rho = {{e^{-\tau}}\o{H(\sqrt{2}-1)}} \to 0 \cr
t_0 =& {{\tau}\o{H}} + {1\o{H}}\log(1-{1 \o \sqrt{2}}) \to +\infty
\cr} \eqn\otra
$$
We see that in the unstable regime (string I for $t_0 \to +\infty$),
 the comoving string
coordinates $(X^1,X^2)$ stay constant whereas the proper string
size $S_-$ and the energy $E_-$  blow up [see eqs.\masin ].
In this regime $\tau$ is proportional to the conformal time
$\eta$. On the other hand, in the stable regime
(string II for $t_0 \to +\infty$), the comoving string
coordinates $(X^1,X^2)$ vanish and $S_-$ and  $E_-$ keep constant,
[see eqs.\masin ]. In this stable regime, $\tau$ is proportional to
the cosmic time $t_0$.
Notice that these results confirm the asymptotic behavior \otra\
discussed in previous works[\desi, \nos , \vene ].

Let us now discuss the solution $q_+(\s,\tau)$ [eq.\solIII ].
Here $\tau$ is a single value function of $q_0$, and hence this
solution describes only one string.

There are two phases here:

(i) $q_0 < 0$ i. e. $\tau < 0$ : contraction phase, $R(\tau)$ decreases,

(ii)$q_0 > 0$ i. e.  $\tau > 0$ : expansion phase, $R(\tau)$ grows.

Here,
$$
q_0\buildrel{\tau \to \pm\infty}\over= \pm {1\o2}(1 - {1\o\sqrt2})
e^{\pm\tau} \to \pm \infty .
$$
The string size and energy are here
$$\eqalign{
S_+(\tau) =& {1\o{\sqrt{2}H}}\tanh\left|{1\o{\sqrt{2}}}\tau\right| \cr
E_+(\tau) =& {1 \o{\a'H}} \left[ 1 - {1 \o{\cosh(\sqrt2 \tau)
+1 - {1\o{\sqrt2}}\sinh(\sqrt2 \tau) }}\right] \cr}
                                        \eqn\tallaII
$$
Therefore, the string contracts from a fixed size $S_+ = {1\o{\sqrt{2} H}}$
and energy $E_+ = {1 \o {\a' H}}$
at $q_0 = -\infty$ during (i) until the collapse at $q_0 = 0$
where $S_+$ vanishes but not the energy which takes the value
$E_+ = {1 \o {2 \a' H}}$ . At this
point the de Sitter space has a minimun size  ${1\o{H}}$. For $q_0 > 0$
, the string size grows from zero  until it takes the
value $S_+ = {1\o{\sqrt{2}H}}$ for $q_0 \to \infty$,
and the energy reaches again the value
 $E_+ = {1 \o {\a' H}}$ , while the de Sitter
 space radius tends to infinity as
$$
R_+(q_0) \buildrel{q_0\to\infty}\over= \; {{q_0}\o H}
\buildrel{q_0\to\infty}\over= (1 - {1 \o {\sqrt{2}}})
{1\o{2 H}}e^\tau
			\eqn\rgrande
$$
This behaviour is different from $q_-(\s,\tau)$ and was not
found before.
Additional solutions follow by replacing
$$  \s \to n\s ,\quad   \tau \to n \tau  ,\quad  n \e Z
	\eqn\rulo
$$
in eqs.\solI\ - \solIII\  . In these solutions the string is winded $n$
times around the $q_1$ axis.

Strings propagating in de Sitter spacetime enjoy as conserved
quantities those associated with the O(3,1) rotations on the
hyperboloid \hiperb\ . They can be written as
$$
L_{AB} = \int_0^{2\pi} d\s \left(\; q_A \;{\dot q}_B - q_B \;{\dot q}_A
\; \right) \quad , \quad 0 \leq A, B \leq 3.
$$
where $L_{AB} = - L_{BA} $. For the solutions $q^{(o)} , q_- $
and $q_+ $ , only $L_{01}$ does not vanish, taking the value:
$$
L_{01} = - L_{10} = n \pi .
$$

\refout
\bigskip
\bigskip
\centerline{FIGURE CAPTIONS}
\medskip
\item{Fig.1.}
The time coordinate $q_0(\tau)$ for the solution $q_-(\sigma,\tau)$
as a function of $\tau$. Since two values of $\tau$ correspond to each
value of $q_0$, the solution $q_-(\sigma,\tau)$ describes {\bf two} strings.

\bye